# Gallium phosphide-on-silicon dioxide photonic devices

Katharina Schneider, Pol Welter, Yannick Baumgartner, Herwig Hahn, Lukas Czornomaz and Paul Seidler

*Abstract*—**The development of integrated photonic circuits utilizing gallium phosphide requires a robust, scalable process for fabrication of GaP-on-insulator devices. Here we present the first GaP photonic devices on SiO$_2$. The process exploits direct wafer bonding of a GaP/Al$_x$Ga$_{1-x}$P/GaP heterostructure onto a SiO$_2$-on-Si wafer followed by removal of the GaP substrate and the Al$_x$Ga$_{1-x}$P stop layer. Photonic devices such as grating couplers, waveguides, and ring resonators are patterned by inductively coupled-plasma reactive-ion etching in the top GaP device layer. The peak coupling efficiency of the fabricated grating couplers is as high as -4.8 dB. Optical quality factors of 17000 as well as second- and third-harmonic generation are observed with the ring resonators. Because the large bandgap of GaP provides for low two-photon absorption at telecommunication wavelengths, the high-yield fabrication of GaP-on-insulator photonic devices enabled by this work is especially interesting for applications in nanophotonics, where high quality factors or low mode volumes can produce high electric field intensities. The large bandgap also enables integrated photonic devices operating at visible wavelengths.**

## I. INTRODUCTION

GALLIUM phosphide (GaP) has been an important material in the photonics industry since the 1960s. Together with GaAsP, it has been the basis for a range of light-emitting devices [1] despite the fact that it has an indirect bandgap for the thermodynamically-favored cubic (zinc blende) crystal structure. More recently, a range of nanophotonic devices and phenomena utilizing GaP has been investigated. These include photonic crystal cavities [2] and coupling of these cavities to emitters, such as fluorescent molecules [3] and nitrogen vacancy (NV) centers [4]. Novel spectrometer concepts [5] as well as second harmonic [6], [7] and sum-frequency generation [8] have also been demonstrated with photonic crystal cavities. The latter is made possible by the non-centrosymmetric crystal structure of GaP, which results in a non-vanishing second order susceptibility $\chi^{(2)}$. Microdisk resonators made of GaP have also been coupled to NV centers [9] and have been used to observe cavity optomechanics [10], where the low two-photon absorption at 1550 nm leads to reduced heating compared to Si-devices [11], [12]. Indeed, the large bandgap (2.26 eV) of GaP makes it an attractive material for mitigating such losses for devices operating at near-infrared wavelengths in addition to enabling functionality at visible wavelengths. The relatively high index of refraction of GaP (n > 3.05 for vacuum wavelengths up to 1600 nm) [13], produces strong confinement of the light and small mode volumes.

All of the nanophotonic structures mentioned above were fabricated on top of a sacrificial Al$_x$Ga$_{1-x}$P layer, which was removed in the region under the device by wet etching to exploit the refractive index contrast between GaP and air. This approach is viable only for free-standing structures; photonic devices such as waveguides, grating couplers or ring resonators cannot be realized in this manner. Fabrication of non-freestanding GaP photonic structures has been achieved utilizing liftoff and transfer onto diamond, where the specific aim was coupling to NV centers [14]–[16]. Large-scale wafer-level fabrication is however not practicable with this technique, and as such it is only sufficient for testing of single devices. Most recently, an approach has been demonstrated exploiting GaP films grown on silicon that are subsequently transferred to glass with an adhesive interlayer (SU-8 2002) [17]. Although this represents a step forward, the use of a polymer limits further processing because of thermal stability, and the process does not provide for full integration, for example with electronic circuitry.

As a scalable and manufacturable solution for fully integrated photonic circuits, we present here the first GaP photonic devices fabricated on SiO$_2$. Our approach makes use of direct wafer bonding of a GaP/Al$_x$Ga$_{1-x}$P/GaP layer structure onto a SiO$_2$-on-Si wafer. The GaP substrate and the Al$_x$Ga$_{1-x}$P stop layer are sequentially removed with selective dry- and wet-etch processes, leading to the desired GaP-on-insulator (GaP-o-I) wafer. A second dry-etch process developed for anisotropic etching of high aspect-ratio structures is then employed to pattern devices in the GaP top layer. Exploiting the significant optical confinement produced by the refractive index contrast between GaP and SiO$_2$ ($n_{GaP} = 3.05$, $n_{SiO_2} = $

The authors are with IBM Research – Zurich, Säumerstrasse 4, CH-8803 Rüschlikon, Switzerland, (e-mail: ksc@zurich.ibm.com and pfs@zurich.ibm.com).

1.44 at 1550 nm), we demonstrate waveguides, low-loss grating couplers (4.8 dB per coupler) and ring resonators with optical quality factors up to 17000. With the latter devices, second- and third-harmonic generation is observed.

## II. DEVICE FABRICATION

The overall process flow is illustrated schematically in Fig. 1. The process begins with the fabrication of a GaP-o-I wafer (steps 1–5). Device structures are subsequently patterned in the top GaP layer with an optimized dry etch process (step 6). The process steps are described in detail in the following sections.

### A. GaP-o-I wafer

First, a GaP/Al$_x$Ga$_{1-x}$P/GaP heterostructure was epitaxially grown on a 2-inch, [100]-oriented, single-side polished, nominally undoped, 400 µm-thick, GaP substrate by metal-organic chemical vapor deposition (MOCVD). The initially deposited 100 nm-thick homoepitaxial GaP buffer layer provided for facile nucleation of the subsequent Al$_x$Ga$_{1-x}$P etch-stop layer. The Al$_x$Ga$_{1-x}$P layer, also 100 nm thick, had an approximate composition of Al$_{0.36}$Ga$_{0.64}$P as determined by X-ray diffraction (XRD). The top GaP device layer was grown with a nominal thickness of 300 nm. All layers were deposited at a susceptor temperature of 650 °C. This layer stack, dubbed the source wafer, was bonded onto a 4-inch Si target wafer capped with 2 µm of SiO$_2$ prepared by thermal dry oxidation at 1050 °C, which eventually becomes the buried oxide of the GaP-o-I wafer [18]. Prior to bonding, the source and target wafers were both coated with approximately 5 nm of Al$_2$O$_3$ by atomic layer deposition (ALD). Megasonic cleaning with ozone-rich deionized water before and after coating with Al$_2$O$_3$ ensured a hydrophilic surface free of organic contamination. The source and target wafer surfaces were brought into intimate contact at room temperature in ambient atmosphere to initiate the bonding followed by annealing at 300 °C for 2 hours to increase the bonding energy.

The original GaP substrate must be removed after bonding. Ideally, a single process which rapidly and selectively removes GaP, stopping on the Al$_x$Ga$_{1-x}$P layer, would be used. To our knowledge, no suitable process for wet etching of GaP selectively with respect to Al$_x$Ga$_{1-x}$P has been published. With the prerequisite of removing several hundred microns in a reasonable time, we tested various solutions for this purpose without success (see Supplemental Material, section I). A selective dry-etch process exists [19] but cannot practically be used alone due to its relatively slow rate and the quantity of material to be removed. We therefore employed a three-step procedure.

First, the GaP substrate was thinned down to ≤ 70 µm by etching in a commercial alkaline solution of K$_3$Fe(CN)$_6$ (Gallium phosphide etchant, Transene). Because {100} faces are polished by this etchant, the peak-to-peak roughness of the initially unpolished back side of the substrate is substantially reduced. The observed etch rate reached a maximum of 670 nm/min for the [100] face, but depended on agitation, temperature, and time in the etchant (presumably because the etchant was being consumed), and was thus hard to control and not sufficiently homogeneous. At this point, the wafer was diced into chips approximately 7 mm × 7 mm. The remaining

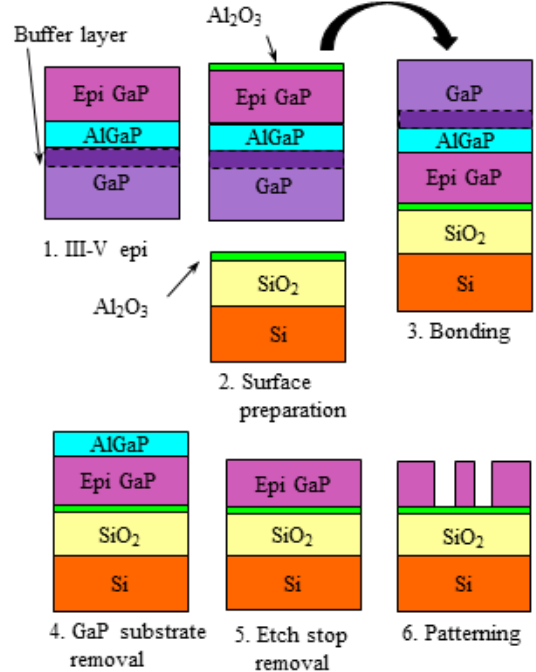

Fig. 1. Schematic of process flow for GaP-o-I device fabrication.

GaP substrate was then further thinned to ≤ 20 µm by inductively-coupled-plasma reactive ion etching (ICP-RIE) with a mixture of H$_2$ and Cl$_2$ that etches homogeneously at a rate of 1.6 µm/min. The process parameters were 20 sccm Cl$_2$, 20 sccm H$_2$, 260 V DC bias, 80 W RF power, 600 W ICP power, 5 mTorr chamber pressure, and 80 °C sample electrode temperature.

Selective etching of GaP in the presence of Al$_x$Ga$_{1-x}$P was provided by the third and final step consisting of another ICP-RIE process, this time with a mixture of Cl$_2$ and CF$_4$. The only previously known selective plasma etch is a SiCl$_4$/SiF$_4$ RIE process developed by Epple et al. [19], with which a selectivity of 126:1 was achieved for GaP with respect to Al$_{0.6}$Ga$_{0.4}$P, a composition with high aluminum content. The etching action is attributed primarily to chlorine-containing species, whereas the fluorine is required for the formation of relatively nonvolatile AlF$_3$, which serves as an etch inhibitor. Because SiF$_4$ is not available in our ICP-RIE tool, we investigated alternative mixtures of various chlorine and fluorine sources, namely Cl$_2$/SF$_6$, Cl$_2$/CF$_4$, Cl$_2$/CHF$_3$, SiCl$_4$/CHF$_3$ and SiCl$_4$/CF$_4$ mixtures (see Supplemental Material, section II). Only Cl$_2$/CF$_4$ and Cl$_2$/CHF$_3$ plasmas exhibited etch rates above 100 nm/min as needed for the removal of several microns within a reasonable timeframe. Optimization of the Cl$_2$/CF$_4$ process (see Supplemental Material, section II for details) gave the following process parameters: 7.5 sccm Cl$_2$, 30 sccm CF$_4$, 240 V DC bias, 100 W ICP power, 60 W RF power, 15 mTorr chamber pressure, and 20 °C sample electrode temperature. GaP was etched with an etch rate of 270 nm/min. The selectivity with respect to Al$_{0.36}$Ga$_{0.64}$P for our bonded heterostructure was estimated to be approximately 120:1. This optimized recipe was used to remove the remaining GaP on top of the Al$_{0.36}$Ga$_{0.64}$P layer.

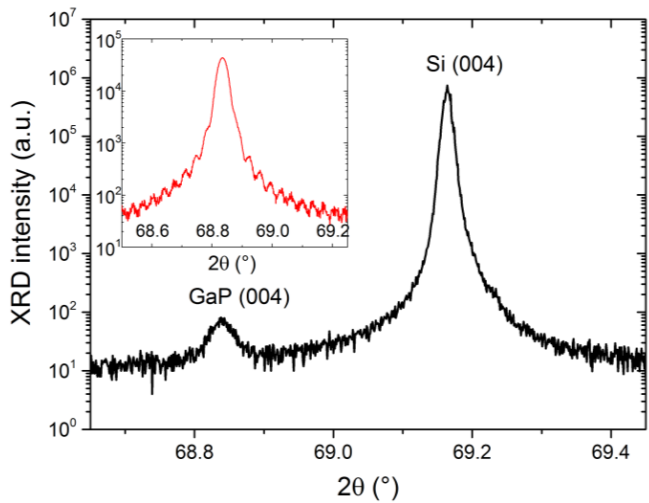

Fig. 3. Omega-2-theta XRD spectrum of a GaP-o-I sample showing diffraction from (004) planes of GaP (left peak) and Si (right peak). The inset displays the GaP diffraction pattern on the same wafer when the diffractometer is aligned to the GaP Bragg reflection to resolve the finite-size oscillations.

Once the original GaP substrate was no longer present, the $Al_{0.36}Ga_{0.64}P$ etch stop layer was easily removed by submerging the sample in concentrated HCl (37 % by weight) for 90 s, resulting in the final GaP-o-I wafer. This step has the opposite selectivity of the previous substrate removal step and leaves the GaP device layer intact. The GaP device layer was characterized by XRD (Fig. 2). A fit of the finite-size oscillations of the GaP (004) Bragg reflection in an omega-2-theta scan yields a GaP layer thickness of 295 nm. The full width at half maximum of the rocking curve for the GaP reflection is 80 arcsec, which corresponds to a threading dislocation density of $2 \times 10^7$ cm$^{-2}$ as calculated from the model of Ayers [20] and indicates that the transferred epitaxial film is of high quality.

### B. GaP patterning

The patterning of devices in the top GaP layer is a critical step. Photonic devices, e.g. grating couplers, waveguides, and ring resonators, require well-defined (typically vertical) and smooth sidewalls to obtain the desired transmission properties, particularly low scattering losses. The fabrication of photonic crystal structures in addition calls for a high-resolution process capable of creating small features with a high aspect ratio and high dimensional accuracy [21], [22]. Most of the studies of anisotropic GaP dry-etching in the literature have focused on increasing the etch rate and maintaining a smooth top-surface morphology [23]–[30]. In these studies, either chlorine-containing species, such as $Cl_2$ or $BCl_3$ [24], [28], [30], [31], or mixtures of $H_2$ and $CH_4$ [25], [27], or a combination of both [29] have been used. Inclusion of Ar and $N_2$, gases which are expected to contribute more to physical as opposed to chemical etching, has also been investigated [24], [28], [30], [32]. The interplay of the various process parameters and gas mixture ratios is complicated, but there are some general trends. Plasma conditions generating a high concentration of chlorine atoms and ions are especially aggressive, with GaP etch rates exceeding 1.5 µm/min [28]. Etching with $BCl_3$ is less aggressive than with $Cl_2$ [28]. The sample electrode temperature has relatively little influence on etch rate and top surface morphology, but does change the sidewall profile, with lower temperatures reducing undercut and roughness, presumably due to passivation effects [29]. For sidewall passivation, $CH_4$ is expected to play an important role [28], but $BCl_3$ and $N_2$ may also be involved [24]. Chemically inert components, such as Ar, that contribute to the etch process in a purely physical manner, tend to increase surface roughness and should be avoided [32].

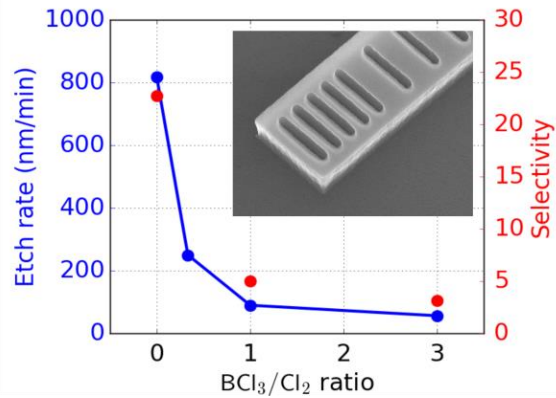

Fig. 2. Dependence of GaP etch rate (blue) and selectivity with respect to HSQ (red) on $BCl_3$-to-$Cl_2$ ratio. Inset: SEM image of a test pattern etched with 17.5 sccm $H_2$, 2.0 sccm $CH_4$, 10.0 sccm $Cl_2$, and 10.0 sccm $BCl_3$ for 2 min.

Keeping the above points in mind, we optimized the dry-etch process in our particular ICP-RIE system (Oxford Instruments PlasmaPro System 100 ICP) starting from a base recipe comprising gas flows of $H_2$ (17.5 sccm), $CH_4$ (2.0 sccm), $Cl_2$ (5.0 sccm), and $BCl_3$ (15.0 sccm), at 600 W ICP-power, 300 V DC bias, 80 °C sample electrode temperature, and 5 mTorr chamber pressure. Optimizing one parameter at a time sequentially, we varied the gas ratios, temperature, ICP power, and chamber pressure. The tests were performed with 4 mm × 6-mm chips diced from [100]-oriented, single-side polished, nominally undoped GaP wafers. A test pattern with various slot openings as small as 50 nm was defined by e-beam lithography (Vistec EBPG 5200ES) using either 4% or 6% hydrogen silsesquioxane (HSQ) in 4-methylpentan-2-one from Dow Corning as a negative resist spin-coated at 6000 rpm, resulting in HSQ film thicknesses of nominally 85 nm and 150 nm, respectively. To improve the adhesion of HSQ to GaP, a thin (3 nm) layer of $SiO_2$ was deposited by ALD prior to spin coating. For the ICP-RIE process, the chips were placed directly on a 100-mm Si carrier wafer backside-cooled with helium, without any affixer such as wax or grease.

As illustrated in Fig. 3, the ratio of $BCl_3$ to $Cl_2$ has a strong influence on the GaP etch rate. For these experiments, the sum of the $BCl_3$ and $Cl_2$ flow rates was kept constant at 20 sccm, and the flow rates of $H_2$ and $CH_4$ were left unchanged. The dramatic decrease in etch rate with added $BCl_3$ might be due to surface passivation by $BCl_3$ [33] or to a reduced density of reactive chlorine-containing species in the plasma, or both. The observation of more sidewall roughness and increased undercutting for decreasing $BCl_3$-to-$Cl_2$ ratio (not shown) points however to a surface passivation role for $BCl_3$. An equal flow of $BCl_3$ and $Cl_2$ yields a reasonable compromise between etch rate and sidewall profile (see the scanning electron microscope (SEM) image in Fig. 3), while providing decent selectivity (5:1) with respect to HSQ and was used in the further optimization of the process as described below.

|  | High-power recipe | Low-power recipe |
|---|---|---|
| Process gases |  |  |
| $H_2$ | 17.5 sccm | 14.5 sccm |
| $CH_4$ | 2.0 sccm | 5.0 sccm |
| $Cl_2$ | 10.0 sccm | 10.0 sccm |
| $BCl_3$ | 10.0 sccm | 10.0 sccm |
| RF Power | 80 W | 120 W |
| DC Bias | 315 V | 390 V |
| ICP Power | 600 W | 200 W |
| Chamber pressure | 5 mTorr | 5 mTorr |
| Temperature | 20°C | 20°C |
| Etch rate | 80 nm/min | 225 nm/min |
| Selectivity with respect to HSQ | 4:1 | 11:1 |

Consistent with the work of Shul et al. [29], we also observed that the extent of undercut and sidewall roughness could be reduced by lowering the temperature, and at a sample electrode temperature of 20 °C, the remaining undercut could be eliminated (inset of Fig. 4(a)). The resultant recipe is referred to as the "high-power" recipe (table 1). Further improvement of the sidewall roughness is expected to be achieved by going to even lower temperatures and adjusting the $BCl_3$-to-$Cl_2$ ratio simultaneously to maintain vertical sidewalls. For prolonged etches, the structures exhibit poorer sidewall roughness even at 20 °C, presumably due to heating of the sample. Periodically interrupting the plasma for 60 s to allow the sample to cool was found to be beneficial but left traces of the process cycling in the sidewall morphology (Fig. 4(a)). One could also consider affixing the sample to the carrier wafer with a grease or other substance to improve thermal contact.

While the external sidewalls of the test structures appeared to be nearly vertical when etched with the high-power recipe, the profile inside small, high-aspect ratio openings was less satisfying. SEM images of cross-sections prepared by focused ion beam (FIB) milling with Ga ions through a test structure (Fig. 4(b)) revealed bowed sidewalls in the slots, where the effect is more pronounced in narrower openings. Such profile issues are common when etching narrow structures and may be caused by accumulation of negative charge near the mouths of the openings [34], leading to deflection of the impinging positive ions towards the inner sidewalls. In addition, we also observed a reduced etch rate in high-aspect ratio openings. For nominally 50 nm-wide slots, we measured an etch rate of 30 nm/min compared to 80 nm/min outside the structure. We can compensate for this well-known phenomenon, commonly referred to as RIE lag [35], by increasing etch time, but the accompanying reduction in selectivity with respect to removal of the e-beam resist must be kept in mind. Even without the RIE lag, the selectivity with respect to HSQ is only 4:1.

To address the need for higher selectivity when etching small openings, the ICP-RIE process was further modified by adjusting the ICP power. The dependence of etch rate and selectivity on ICP power, with all other parameters of the high-power recipe left unchanged, is shown in Fig. 5. While the GaP etch rate is essentially constant from 400 W to 600 W, we observe a significant rise in the etch rate at 200 W, a somewhat counterintuitive result given that increased ICP power is associated with a higher plasma density, which typically would increase the chemical component of etching. Several mechanisms could account for this behavior, such as a change in the plasma chemistry at higher ICP powers, leading to either creation of species that are effective at passivation or removal of species responsible for etching. For the latter possibility, sputter desorption of etchants from the surface may play a role [28].

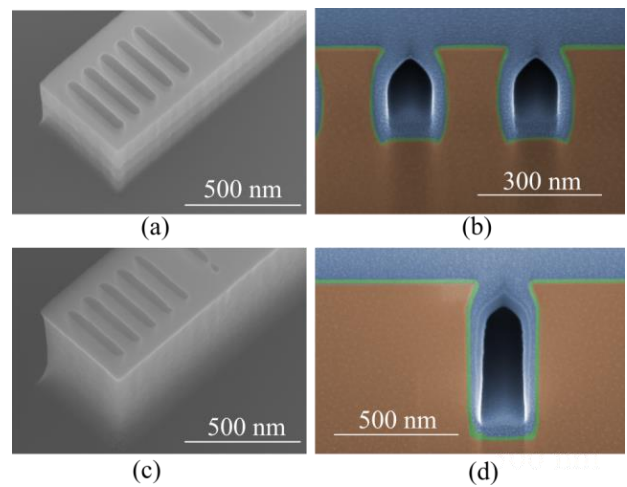

Fig. 4. (a) and (c) SEM images of test structures etched with the high-power recipe and low-power recipe, respectively. (b) and (d) False-color SEM images at 52° tilt angle of cross-sections prepared by FIB of these test structures. The GaP (red) is covered by a thin $SiO_2$ layer (green), which was deposited for better contrast, and a platinum layer (blue), which was deposited to protect the devices during FIB milling. The dark regions are voids. The design width of the slots is 150 nm.

One other parameter that could influence etch performance is chamber pressure. We found that an increase from 5 mTorr to 10 mTorr roughened the sidewalls and led to undercutting of the patterned structures. At lower pressure, the plasma was not stable. Therefore, a pressure of 5 mTorr was maintained.

Finally, taking advantage of the known passivation behavior of $CH_4$ [29], the $H_2$-to-$CH_4$ ratio was reduced from the usual value of 8.75:1 to 2.9:1 in order to increase the anisotropy when etching with an ICP power of 200 W. The resulting "low-power" recipe is specified in table 1. An SEM image of a test pattern fabricated with the low-power recipe (Fig. 4(c)) indicates smooth, nearly vertical etching of the outer sidewalls for features etched over 800 nm deep. Owing to the reduced ICP power, periodic interruption of the etching to allow cooling of the sample was no longer necessary. The pronounced foot that forms at the bottom of the structures may be a consequence of the surface passivation process, as it can be reduced by cycling with an oxygen plasma. It is absent on GaP-o-I substrates when the GaP is etched through to the underlying oxide (Fig. 6). The higher selectivity with respect to HSQ (11:1) permits the use of thinner resist layers, which should improve dimensional accuracy. Except for very small openings (e.g. 50-nm width), we observe a better sidewall profile than for the high-power recipe, as illustrated by the SEM image in Fig. 4(d) of a cross-section prepared by FIB, where there is significant deviation from verticality only at the top of the opening.

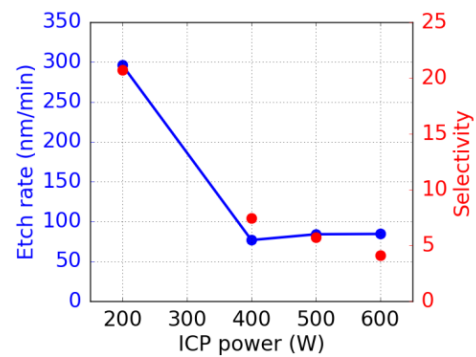

Fig. 5. Dependence of GaP etch rate (blue) and selectivity with respect to HSQ (red) on ICP power.

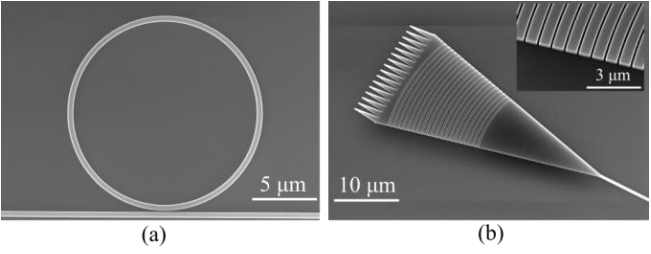

(a)                        (b)

Fig. 6. SEM images of fabricated photonic devices. (a) Ring resonator comprising a 400 nm-wide circular waveguide with a radius of 7.5 µm and the associated 400 nm-wide bus waveguide. The gap between the ring and the bus waveguide is 160 nm. (b) Grating coupler at 30° tilt angle.

### C. Devices

A series of device structures was fabricated on a 6 mm × 7 mm GaP-o-I chip. The structures included 400 nm-wide ridge waveguides with lengths up to 2.5 mm as well as circular ring resonators, also 400 nm wide, with radii ranging from 5 µm to 15 µm and coupling gaps to the associated bus waveguide ranging from 80 nm to 240 nm (Fig. 6(a)). Focusing grating couplers located at the ends of the waveguides enabled local testing of the devices without end-facet polishing. The focusing grating-couplers (Fig. 6(b)) consisted of 19 curved lines of GaP and were designed with either a periodicity of 764 nm and a duty cycle of 80% or a periodicity of 728 nm and a duty cycle of 85%. The two designs have similar transmission performance, and in both cases, the couplers have a length of 35 µm and are designed for nearly vertical coupling (10° at 1550 nm). Sub-wavelength-sized wedges are incorporated opposite the waveguide to reduce reflections [36], [37].

The structures were defined by e-beam lithography using a pre-coating of 3 nm of $SiO_2$ deposited by ALD and 6% HSQ as resist. Pattern transfer was carried out with the low-power ICP-RIE recipe, as described above in section II-B. The waveguides, ring resonators, and grating-couplers were all fabricated in the same process step and were fully etched through to the underlying $SiO_2$ layer. After etching, the chip was exposed to an oxygen plasma (600 W) for 3 min. This step serves to prevent the formation of droplet-shaped residues otherwise observed on the GaP sidewalls. Finally, the HSQ was removed by submerging the chip in standard buffered oxide etchant for 10 s.

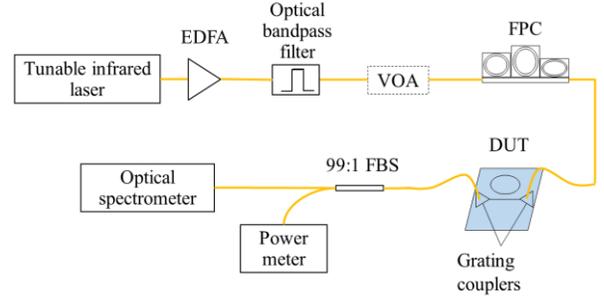

Fig. 7. Measurement apparatus used for transmission measurements.

### III. OPTICAL CHARACTERIZATION

The devices were characterized by means of their optical transmission behavior. A schematic of the apparatus is shown in Fig. 7. All measurements were performed in air at atmospheric pressure with the chip resting on an aluminum block held at 20.0 °C, as measured with an integrated thermistor, and controlled with a Peltier element. Continuous-wave infrared light from a tunable external-cavity laser (Photonetics Tunics-Plus) was directed through a cleaved single-mode optical fiber into the input grating coupler of the device under test (DUT). A fiber polarization controller (FPC) was used to align the polarization of the light with the TE design orientation of the grating couplers. Light emitted from the output grating coupler on the other side of the device was collected with another cleaved single-mode optical fiber, which was connected to a power meter (EXFO IQ 1600), either directly or through a fiber beam splitter (FBS), for monitoring and recording of transmission spectra. For measurements of second and third harmonic generation, an optical spectrometer (Ocean Optics USB2000) was used, where the second cleaved fiber was positioned either over the output grating coupler or directly above the ring resonator to collect the scattered light. The latter configuration was necessary for observation of the third harmonic, as it is absorbed by the GaP, and in any case, the grating couplers were not designed for the harmonic wavelengths. For characterization of the power dependence of harmonic generation, an erbium-doped fiber amplifier (EDFA), a bandpass filter, and a variable optical attenuator (VOA) were introduced between the tunable laser and the FPC.

The grating couplers exhibited a Gaussian-shaped transmission profile (Fig. 8(a)) that was red shifted with respect to the design wavelength of 1550 nm presumably because of deviations from design dimensions in the fabricated structures. However, when the input and output optical fibers were aligned at an angle of 20° to the chip normal instead of the 10° design angle, the center wavelength was at 1553 nm and the efficiency per coupler was -4.8 dB. In principle, the losses could differ between the input and output grating couplers; we have assumed that they are identical and taken half of the total loss occurring between the two cleaved fibers. For comparison, silicon focusing grating couplers with a design similar to that described here, i.e. fully etched with no cladding, have more than 10 dB loss per coupler [11], presumably to some extent because of the higher index contrast. Highly optimized, partially etched and cladded focusing grating couplers in silicon can however have a coupling efficiency approaching -1 dB [38]. With the same techniques, it should be possible to improve the transmission of the GaP grating couplers.

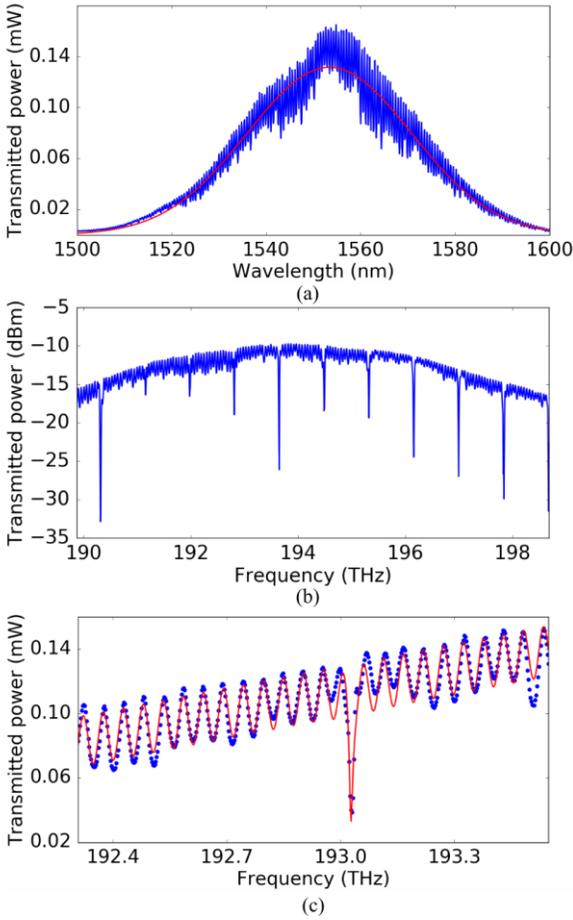

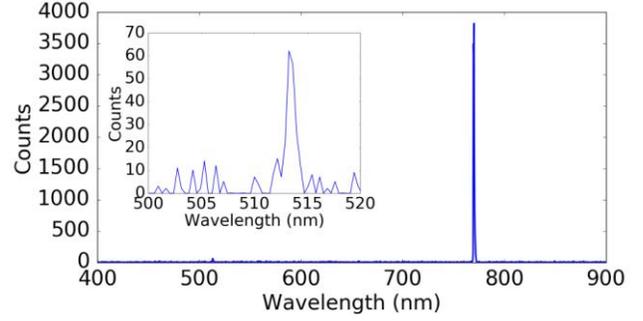

Fig. 9. Spectrum of the diffracted light collected with the output fiber positioned above a ring resonator with 3.7 mW entering the device and an integration time of 5 s. The peak at 770 nm is the SHG signal. At 513 nm (inset), the THG signal is observed.

Fig. 8. (a) Transmission through two grating couplers with a period of 764 nm and a duty cycle of nominally 80% connected by a 550 µm-long waveguide measured with 1.17 mW input power and the input and output optical fibers at an angle of 20° to normal. The maximum coupling efficiency is -4.8 dB as obtained by fitting a Gaussian function (red). (b) Transmission spectrum of a ring resonator with a radius of 15 µm. (c) Transmission spectrum of a device with a radius of 12.5 µm near one particular resonance with quality factor of $Q = 17000 \pm 2000$. The oscillating background is due to reflections between the grating couplers.

For waveguide lengths up to 2200 µm, variations in the grating coupler losses dominate over losses in the waveguides. In other words, the propagation loss is not significant compared to the scatter in the measured transmission and would require significantly longer waveguides to be determined. From the Q factors observed for the ring resonators (see below) we can however estimate an upper bound of 10 dB/cm for the propagation loss.

Fig. 8(b) displays the transmission spectrum of a ring resonator comprising a 400 nm-wide circular waveguide with a radius $r = 15$ µm evanescently coupled to a 400 nm-wide bus waveguide through a 240-nm coupling gap. The observation of a single family of resonances is consistent with the small cross-section of the waveguide supporting only the fundamental TE-polarized mode. We attribute the splitting of some of the resonances into double resonances (not visible in Fig. 8) to coupling with counter propagating modes excited by backscattering from imperfections in the waveguide such as surface roughness. The free spectral range between adjacent resonances is given by $\Delta_{FSR} = c/n_g L$, where $n_g = n_{eff} + \nu \frac{dn_{eff}}{d\nu}$ is the group refractive index, $n_{eff}$ is the effective refractive index, $\nu$ is the frequency, $c$ is the speed of light and $L = 2\pi r$ is the length of the resonator. From finite element simulations at $\lambda = 1550$ nm, $n_{eff} = 1.9375$ and $n_g = 4.00$ can be inferred, which leads to $\Delta_{FSR} = 0.80$ THz (equivalent to 6.3 nm at this wavelength). The measured free spectral range for the device in Fig. 8(b) is 0.83 THz.

Fitting a Lorentzian function to a resonance, as shown in Fig. 8(c), allows one to calculate the quality factor $Q = \nu_0/\Delta\nu$, where $\nu_0$ is the resonance frequency and $\Delta\nu$ is the full width at half maximum of the resonance dip. The periodic background ripple is due to interference between reflections from the grating couplers and can be modelled with an Airy function, which is the typical description of the sum of the longitudinal mode profiles of a Fabry-Perot cavity [39]. The best measured devices have loaded quality factors of $Q = 17000$.

The non-centrosymmetric crystal structure of GaP implies that it has a non-vanishing second-order susceptibility ($\chi^{(2)}$), which leads to nonlinear optical effects, e.g. second-harmonic generation (SHG) [40]. A spectrum of the light gathered with the output fiber positioned above a ring resonator while pumping at 1539 nm with 3.7 mW injected into the device (i.e. after the input grating coupler) is shown in Fig. 9. In addition to the strong SHG signal observed at 770 nm, third-harmonic generation (THG) is detected around 513 nm (inset in Fig. 9). Although the THG is much weaker than the SHG, it is in the green portion of the visible spectrum and sufficiently intense to be easily seen with the naked eye. SHG has been previously reported at low input powers in two-dimensional photonic-crystal cavities and waveguides made of GaP [6], [41]. To our knowledge this is the first report of third harmonic generation in a GaP photonic crystal cavity, however observation of weak THG in GaP microdisks has been reported [42].

The dependence of the second-harmonic intensity on the input power at the fundamental wavelength for a resonance positioned at 1537.67 nm is shown in Fig. 10(a). A quadratic fit of the data confirms the second-order nature of the process. The dependence of the second harmonic intensity on the input wavelength is shown in Fig. 10(b). The SHG signal was detected with the cleaved fiber positioned above the ring resonator. The square of the inverted transmitted power (i.e. the power dropped into the ring) measured separately is overlaid on the plot. The peak of the second harmonic light is slightly blue-shifted with respect to the resonance at the pump wavelength, which can be attributed to dispersion and the consequent misalignment of the resonances at the second harmonic frequency with respect to those at the fundamental frequency [43].

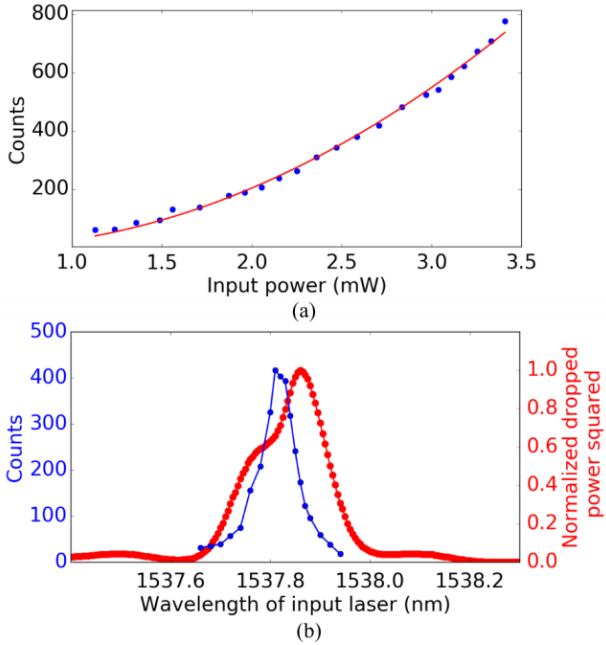

Fig. 10. (a) Intensity of the second harmonic as a function of the power entering the bus waveguide of the ring resonator. A quadratic fit (solid line) matches well the measured data. (b) Second harmonic intensity as a function of input wavelength (blue) as measured with the output fiber above the ring resonator. A transmission measurement at the fundamental wavelength has been inverted, squared and rescaled (red).

## IV. Conclusion

In this paper, we have described a complete process flow for the fabrication of the first GaP-on-SiO$_2$ photonic devices. Several new process steps have been demonstrated: direct wafer bonding of a GaP/Al$_x$Ga$_{1-x}$P/GaP heterostructure to a SiO$_2$-on-Si wafer; a GaP substrate removal procedure involving a combination of wet and dry etching, where the key step is a final Cl$_2$/CF$_4$ ICP-RIE process that stops selectively on the Al$_x$Ga$_{1-x}$P sacrificial layer; and two alternative ICP-RIE recipes employing H$_2$/CH$_4$/Cl$_2$/BCl$_3$ plasmas that allow device patterning of high aspect-ratio devices with relatively smooth, nearly vertical, outer sidewalls. Simple building blocks for integrated photonic circuits have been realized, such as waveguides, grating couplers with losses below 5 dB, and ring resonators with quality factors of 17000. Intense second-harmonic as well as third-harmonic light can be generated with the ring resonators.

The ability to work with full GaP-on-SiO$_2$ wafers makes our approach inherently higher yield and more versatile than the liftoff-and-transfer method previously employed for GaP-on-diamond devices [15], [16]. It is also more robust with respect to further processing and allows more direct integration than bonding methods employing polymers [17]. Indeed, we expect this GaP-o-I technology platform to enable both new device architectures for GaP and their integration into photonic and electronic circuits. For example, the various device concepts utilizing photonic crystal structures, which to-date have been demonstrated as isolated, free-standing devices interrogable only with free-space optics such as high-numerical aperture microscope objectives or tapered fibers, could instead be connected directly with other on-chip elements via waveguides. Devices operating in the visible spectrum, for example, spectrometers or structures which couple to atoms [44], quantum dots or NV centers [16], could be similarly integrated. In the area of nonlinear optics, ring resonators could be designed with a second bus waveguide for efficient collection and out-coupling of the SHG signal [43], and, eventually, on-chip frequency conversion of integrated emitters could be explored. Alternatively, the observation of THG suggests that creation of a Kerr frequency comb may be possible [44]. These are just a few of the numerous opportunities created for GaP nanophotonics.

## Acknowledgment

We gratefully acknowledge the contributions of Antonis Olziersky to the e-beam lithography, Stefan Abel to the design of the grating couplers, Daniele Caimi for the bonding of the wafers, Ute Drechsler for assistance with the ICP-RIE, and Dalziel Wilson for discussions of some of the measurements. This work was supported by the European Union's Horizon 2020 Programme for Research and Innovation under grant agreement No 732894 (FET Proactive HOT).

## References

[1] E. F. Schubert, "History of light-emitting diodes" in *Light-emitting diodes*, 2nd ed. Cambridge: Cambridge University Press, 2006, pp.1-26.

[2] K. Rivoire, A. Faraon, and J. Vučković, "Gallium phosphide photonic crystal nanocavities in the visible," *Appl. Phys. Lett.*, vol. 93, no. 6, 2008, Art. no. 063103.

[3] K. Rivoire *et al.*, "Lithographic positioning of fluorescent molecules on high- Q photonic crystal cavities," *Appl. Phys. Lett.*, vol. 95, no. 12, 2009, Art. no. 123113.

[4] L. Li, T. Schröder, E. H. Chen, H. Bakhru, and D. Englund, "One-dimensional photonic crystal cavities in single-crystal diamond," *Photonics Nanostructures - Fundam. Appl.*, vol. 15, pp. 130–136, 2015.

[5] X. Gan, N. Pervez, I. Kymissis, F. Hatami, and D. Englund, "A high-resolution spectrometer based on a compact planar two dimensional photonic crystal cavity array," *Appl. Phys. Lett.*, vol. 100, no. 23, 2012, Art. no. 231104.

[6] K. Rivoire, Z. Lin, F. Hatami, W. T. Masselink, and J. Vučković, "Second harmonic generation in gallium phosphide photonic crystal nanocavities with ultralow continuous wave pump power," *Opt. Express*, vol. 17, no. 25, pp. 22609–22615, 2009.

[7] G. Shambat, K. Rivoire, J. Lu, F. Hatami, and J. Vučković, "Tunable-wavelength second harmonic generation from GaP photonic crystal cavities coupled to fiber tapers," *Opt. Express*, vol. 18, no. 12, pp. 12176–84, 2010.

[8] K. Rivoire, Z. Lin, F. Hatami, and J. Vučković, "Sum-frequency generation in doubly resonant GaP photonic crystal nanocavities," *Appl. Phys. Lett.*, vol. 97, no. 4, 2010, Art. no. 043103.

[9] C. Santori, P. E. Barclay, K.-M. C. Fu, R. G. Beausoleil, S. Spillan, and M. Fisch, "Nanophotonics for quantum optics using nitrogen-vacancy centers in diamond," *Nanotechnology*, vol. 21, no. 27, p. 274008, 2010.

[10] M. Mitchell, A. C. Hryciw, and P. E. Barclay, "Cavity optomechanics

in gallium phosphide microdisks," *Appl. Phys. Lett.*, vol. 104, no. 14, 2014, Art. no. 141104.

[11] P. Seidler, K. Lister, U. Drechsler, H. Rothuizen, J. Hofrichter, and T. Stöferle, "Photonic crystal nanobeam cavities with an ultrahigh quality factor-to-modal volume ratio," *Opt. Express*, vol. 21, no. 26, pp. 32468–32483, 2013.

[12] R. Riedinger *et al.*, "Non-classical correlations between single photons and phonons from a mechanical oscillator," *Nature*, vol. 530, no. 7590, pp. 313–316, 2016.

[13] W. L. Bond, "Measurement of the refractive indices of several crystals," *J. Appl. Phys.*, vol. 36, pp. 1674–1677, 1965.

[14] P. E. Barclay, K.-M. Fu, C. Santori, and R. G. Beausoleil, "Hybrid photonic crystal cavity and waveguide for coupling to diamond NV-centers," *Optics*, vol. 17, no. 12, pp. 9588–9601, 2009.

[15] P. E. Barclay, K. M. C. Fu, C. Santori, and R. G. Beausoleil, "Chip-based microcavities coupled to nitrogen-vacancy centers in single crystal diamond," *Appl. Phys. Lett.*, vol. 95, no. 19, 2009, Art. no. 191115.

[16] N. Thomas, R. J. Barbour, Y. C. Song, M. L. Lee, and K. M. C. Fu, "Waveguide-integrated single-crystalline GaP resonators on diamond," *Opt. Express*, vol. 22, no. 11, pp. 13555–13564, 2014.

[17] H. Emmer *et al.*, "Fabrication of Single Crystal Gallium Phosphide Thin Films on Glass," *Sci. Rep.*, vol. 7, no. 1, p. 4643, 2017.

[18] L. Czornomaz *et al.*, "Scalability of Ultra-thin-body and BOX InGaAs MOSFETs on Silicon," in *Proc. of the European Solid-State Device Research Conference (ESSDERC)*, Bucharest, Romania, 2013, no. 6, pp. 143–146.

[19] J. H. Epple, C. Sanchez, T. Chung, K. Y. Cheng, and K. C. Hsieh, "Dry etching of GaP with emphasis on selective etching over AlGaP," *J. Vac. Sci. Technol. B Microelectron. Nanom. Struct. Process. Meas. Phenom.*, vol. 20, no. September, p. 2252, 2002.

[20] J. E. Ayers, "The measurement of threading dislocation densities in semiconductor crystals by X-ray diffraction," *J. Cryst. Growth*, vol. 135, no. 1–2, pp. 71–77, Jan. 1994.

[21] K. Schneider and P. Seidler, "Strong optomechanical coupling in a slotted photonic crystal nanobeam cavity with an ultrahigh quality factor-to-mode volume ratio," *Opt. Express*, vol. 24, no. 13, pp. 13850–13865, 2016.

[22] P. Seidler, "Optimized process for fabrication of free-standing silicon nanophotonic devices," *J. Vac. Sci. Technol. B, Nanotechnol. Microelectron. Mater. Process. Meas. Phenom.*, vol. 35, no. 3, p. 31209, 2017.

[23] J. W. Lee, C. J. Santana, C. R. Abernathy, S. J. Pearton, and F. Ren, "Wet chemical etch solutions for AlxGa1-xP," *J. Electrochem. Soc.*, vol. 143, no. 1, pp. L1–L3, 1996.

[24] J. W. Lee *et al.*, "Plasma Etching of III-V Semiconductors in BCl3 Chemistries: Part I: GaAs and Related Compounds," *Plasma Chem. Plasma Process.*, vol. 17, no. 2, pp. 155–167, 1997.

[25] S. J. Pearton *et al.*, "Dry etching characteristics of III-V semiconductors in microwave BCl3 discharges," *Plasma Chem. Plasma Process.*, vol. 13, no. 2, pp. 311–332, 1993.

[26] S. J. Pearton *et al.*, "High microwave power electron cyclotron resonance etching of III–V semiconductors in CH4/H2/Ar," *J. Vac. Sci. Technol. B Microelectron. Nanom. Struct.*, vol. 14, no. 1, p. 118, 1996.

[27] S. Pearton, J. Lee, E. Lambers, and C. Abernathy, "Comparison of Dry Etching Techniques for III-V Semiconductors in CH 4/H 2/Ar Plasmas," *J. Electrochem. Soc.*, vol. 143, no. 2, pp. 752–758, 1996.

[28] R. J. Shul *et al.*, "High-density plasma etching of compound semiconductors," *J. Vac. Sci. Technol. A Vacuum, Surfaces, Film.*, vol. 15, no. 3, p. 633, 1997.

[29] R. J. Shul, A. J. Howard, C. B. Vartuli, P. A. Barnes, and W. Seng, "Temperature dependent electron cyclotron resonance etching of InP, GaP, and GaAs," *J. Vac. Sci. Technol. A Vacuum, Surfaces Film.*, vol. 14, no. 3, pp. 1102–1106, 1996.

[30] J. W. Lee, J. Hong, E. S. Lambers, C. R. Abernathy, and S. J. Pearton, "C12-Based Dry Etching of GaAs, AlGaAs, and GaP," *J. Electrochem. Soc.*, vol. 143, no. 6, pp. 2010–2014, 2010.

[31] G. Smolinsky, R. P. Chang, and T. M. Mayer, "Plasma etching of III–V compound semiconductor materials and their oxides," *J. Vac. Sci. Technol.*, vol. 18, no. 1, p. 12, 1981.

[32] S. H. Yang and P. R. Bandaru, "An experimental study of the reactive ion etching (RIE) of GaP using BCl3 plasma processing," *Mater. Sci. Eng. B Solid-State Mater. Adv. Technol.*, vol. 143, no. 1–3, pp. 27–30, 2007.

[33] D. L. Flamm and V. M. Donnelly, "The Design of Plasma Etchants," *Plasma Chem. Plasma Process.*, vol. 1, no. 4, pp. 317–363, 1981.

[34] C. Welch, "Nanoscale Etching in Inductively Coupled Plasmas," Oxford Instruments, Oxford, United Kingdom, 2011. [Online]. Available: http://www.oxfordplasma.de/pla_news/wh_paper/OIPT - Nanoscale Etching of ICP Plasmas 2011.pdf. [Accessed on: May 3, 2017].

[35] H. Jansen *et al.*, "RIE lag in high aspect ratio trench etching of silicon," *Microelectron. Eng.*, vol. 35, no. 1–4, pp. 45–50, 1997.

[36] R. Waldhausl, B. Schnabel, E.-B. Kley, and A. Brauer, "Efficient focusing polymer waveguide grating couplers," *Electron. Lett.*, vol. 33, no. 7, pp. 623–624, 1997.

[37] S. Assefa *et al.*, "A 90nm CMOS Intergated Nanophotonics Technology for 25Gbps WDM Optical Communications Applications," in *Proc. of 2012 IEEE International Electron Devices Meeting (IEDM)*, San Francisco, CA, 2012, pp. 809–811.

[38] A. Mekis *et al.*, "A grating-coupler-enabled CMOS photonics platform," *IEEE J. Sel. Top. Quantum Electron.*, vol. 17, no. 3, pp. 597–608, 2011.

[39] N. Ismail, C. C. Kores, D. Geskus, and M. Pollnau, "Fabry-Pérot resonator: spectral line shapes, generic and related Airy distributions, linewidths, finesses, and performance at low or frequency-dependent reflectivity," *Opt. Express*, vol. 24, no. 15, p. 16366, 2016.

[40] B. E. A. Teich and M. C. Saleh, "Nonlinear optics," in *Fundamentals of Photonics*, New York: Wiley, 1992, pp. 873–917.

[41] K. Rivoire, S. Buckley, F. Hatami, and J. Vučković, "Second harmonic generation in GaP photonic crystal waveguides," *Appl. Phys. Lett.*, vol. 98, no. 26, 2011, Art. no. 263113.

[42] D. P. Lake, M. Mitchell, H. Jayakumar, L. F. Dos Santos, D. Curic, and P. E. Barclay, "Efficient telecom to visible wavelength conversion in doubly resonant gallium phosphide microdisks," *Appl. Phys. Lett.*, vol. 108, no. 3, 2016, Art. no. 031109.

[43] X. Guo, C. Zou, and H. Tang, "Second-harmonic generation in aluminum nitride microrings with 2500 %/ W conversion efficiency," *Optica*, vol. 3, no. 10, p. 1126, 2016.

[44] A. González-Tudela, C.-L. Hung, D. E. Chang, J. I. Cirac, and H. J. Kimble, "Subwavelength vacuum lattices and atom–atom interactions in two-dimensional photonic crystals," *Nat. Photonics*, vol. 9, no. 5, pp. 320–325, 2015.

[45] M. Pu, L. Ottaviano, E. Semenova, and K. Yvind, "Efficient frequency comb generation in AlGaAs-on-insulator," *Optica*, vol. 3, no. 8, pp. 8–11, 2016.

# Gallium phosphide-on-silicon dioxide photonic devices – Supplemental Material

Katharina Schneider, Pol Welter, Yannick Baumgartner, Herwig Hahn, Lukas Czornomaz and Paul Seidler

## V. WET ETCHING OF GaP VERSUS $Al_xGa_{1-x}P$

A description of the various solutions tested for selective etching of GaP in the presence of $Al_xGa_{1-x}P$ and their etching behavior are listed in Table I and Table II. Tests were performed with 4-mm × 4-mm, [100]-oriented, GaP and $Al_{0.18}Ga_{0.82}P$ chips with a 50 nm-thick $SiO_2$ mask. Tests on $Al_{0.18}Ga_{0.82}P$ were conducted only if the etching of GaP was sufficiently fast and uniform. In no case was adequate selectivity achieved.

TABLE I
ETCH RATES OF GaP AND $Al_{0.18}Ga_{0.82}P$ FOR VARIOUS ETCHANT SOLUTIONS

| Etchant mixtures by volume[a] | Hotplate temperature [°C] | Etch rate of GaP [nm/min] | Etch rate of $Al_{0.18}Ga_{0.82}P$ [nm/min] |
|---|---|---|---|
| Citric acid/$H_2O_2$ (1:5) | | < 0.5 | |
| Citric acid/$H_2O_2$ (1:1) | | < 0.5 | |
| Citric acid/$H_2O_2$ (5:1) | | ~1.3 | |
| Citric acid/$H_2O_2$ (25:1) | | ~1.3 | |
| Citric acid/$H_2O_2$ (10:1) | 100 | 15 | 13 |
| $H_2O_2$/$NH_4OH$ (1:10) | 80 | 5 | 3 |
| $H_2O_2$/$NH_4OH$ (1:1) | 80 | 7 | 5 |
| $H_2O_2$/$NH_4OH$ (10:1) | 80 | 7 | |
| $H_2O_2$/$NH_4OH$ (100:1) | 80 | 6 | 5 |
| $H_2O_2$/$NH_4OH$ (10:1) | 100 | 16 | 8 |
| HCl | | 0 | >116 |
| HCl/$H_2O_2$ (1:3) | | ~10[b] | |
| HCl/$H_2O_2$ (1:1) | | >500[b] | |
| HCl/$H_2O_2$ (3:1) | | ≤18000[b] | fast |
| HCl/$H_2O_2$ (10:1) | | >1000[b] | |
| HCl/$HNO_3$ (3:1) | | 210[c] | 276 |
| HCl/$HNO_3$ (3:1) | 60 | 1600[c] | |

The experiments were carried out at room temperature unless otherwise noted, in which case the temperature indicated is that of the hotplate heating the solution.

[a]The concentrations by weight in water of the component solutions are as follows:
Citric acid: 50% citric acid monohydrate
$H_2O_2$: 30–31%
$NH_4OH$: 58–62% (28–30% $NH_3$)
KOH: 44%
HCl: 37%
$HNO_3$: 70%
[b]The high etch rates are estimates, as the etchant is consumed quickly and etching is not uniform.
[c]Circular pits on surface, etching not uniform.

## VI. SELECTIVE ICP-RIE OF GaP IN THE PRESENCE OF $Al_xGa_{1-x}P$

Experiments to develop a selective ICP-RIE process for removal of GaP stopping on $Al_xGa_{1-x}P$ were performed with 4-mm × 4-mm GaP and $Al_{0.18}Ga_{0.82}P$ chips with either a 50 nm- or 100 nm-thick $SiO_2$ mask. A list of the alternative gas mixtures with various sources of chlorine and fluorine species

TABLE II
ETCH RATES OF GaP AND $Al_{0.18}Ga_{0.82}P$ FOR SOLUTIONS OF $K_3Fe(CN)_6$ AND KOH

| $K_3Fe(CN)_6$ concentration [M] | KOH concentration [M] | Etch rate of GaP [nm/min] | Etch rate of $Al_{0.18}Ga_{0.82}P$ [nm/min] |
|---|---|---|---|
| 0.97 | 0.32 | 234 | 550 |
| 0.97 | 0.032 | 100 | 130 |
| 0.90 | 0.01 | 10 | 12 |
| 0.97 | 0.003 | 0 | 2 |

that were investigated and the observed etch rates are reported
The experiments were carried out at room temperature. The etchant polishes {100} surfaces of GaP.

in Table III. The experimental parameters were fixed at 100 W ICP power, 60 W RF power, 15 mTorr chamber pressure, and 20°C sample electrode temperature. Etch tests of $Al_{0.18}Ga_{0.82}P$ were carried out only for those cases where the etch rate of GaP was reasonably high.

TABLE III
ETCH TESTS OF GaP AND $Al_{0.18}Ga_{0.82}P$ WITH VARIOUS SOURCES OF CHLORINE AND FLUORINE SPECIES.

| Cl compound (sccm) | F compound (sccm) | DC bias [V] | Etch rate GaP [nm/min] | Etch rate AlGaP [nm/min] | Selectivity |
|---|---|---|---|---|---|
| $Cl_2$ (10 ) | $SF_6$ (40) | 152 | 4 | – | – |
| $Cl_2$ (20 ) | $SF_6$ (20) | 170 | 8 | – | – |
| $Cl_2$ (7.5 ) | $CF_4$ (30) | 242 | 275 | 8 | 37 |
| $Cl_2$ (7.5 ) | $CHF_3$ (30) | 258 | 127 | 6 | 21 |
| $SiCl_4$ (7.5) | $CF_4$ (30) | 250 | 23 | – | – |
| $SiCl_4$ (7.5) | $CHF_3$ (30) | 250 | 23 | – | – |

As $SF_6$ seems to suppress the etching capabilities of $Cl_2$ quite strongly, and $SiCl_4$ yields only low rates in comparison to those for $Cl_2$, we further optimized the process based on $Cl_2$/$CF_4$ mixtures, the results of which are tabulated in Table IV. Here, the DC bias was targeted at two values, either 180 V or 240 V. Because the plasma erodes the $SiO_2$ mask, etch times were generally limited to 2 min (except as indicated), which of course bounds the achievable accuracy of the etch rates, especially for $Al_{0.18}Ga_{0.82}P$. The observed selectivity falls primarily into two clusters, one group of results at ≲1:1 and one around 30:1. Overall, higher selectivity is achieved for the lower chamber pressure of 15 mTorr. The other parameters seem to have a more complex influence on selectivity, presumably because of a competition between formation and removal of a passivation layer. While more experiments exploring the full parameter space may provide further optimization, the demonstrated selectivity is sufficient to complete the GaP substrate removal of the bonded wafers, and the second recipe in Table IV was used for this purpose.

The dependence on the Al content was investigated by comparing the behavior of $Al_{0.36}Ga_{0.64}P$ and $Al_{0.18}Ga_{0.82}P$ when etched with recipe 2. The selectivity increased by a factor of 1.5 for the higher Al content. When applying this recipe to the bonded $GaP/Al_{0.36}Ga_{0.64}P/GaP$ heterostructure, we observe an even higher selectivity estimated to be around 120. This might be due to a difference in the $Al_{0.36}Ga_{0.64}P$ interface when it is directly capped with GaP during MOCVD growth. Indeed, preliminary experiments suggest that initial etch rates may depend on past exposure of the top surface of the various GaP and $Al_xGa_{1-x}P$ samples.

TABLE IV
ETCH TESTS OF GaP AND $Al_{0.18}Ga_{0.82}P$ WITH A $Cl_2/CF_4$ PLASMA

| Recipe | $Cl_2$ flow (%)[a] | $CF_4$ flow (%)[a] | Pressure [mTorr] | ICP power [W] | DC bias [V] | Etch rate GaP [nm/min] | Etch rate AlGaP [nm/min] | Selectivity |
|---|---|---|---|---|---|---|---|---|
| 1 | 11 | 89 | 15 | 100 | 243 | 0 | 0 | - |
| 2 | 20 | 80 | 15 | 100 | 241 | 257 | 7.5 | 37:1 |
| 3 | 29 | 71 | 15 | 100 | 238 | 320 | 11 | 29:1 |
| 4 | 20 | 80 | 15 | 130 | 242 | 265 | 8.5 | 31:1 |
| 5[b] | 20 | 80 | 15 | 130 | 183 | 208 | 7 | 30:1 |
| 6 | 19 | 81 | 15 | 180 | 185 | 4 | 22 | 0.18:1 |
| 7 | 12 | 88 | 15 | 180 | 244 | 11 | 11 | 1:1 |
| 8 | 25 | 75 | 15 | 180 | 238 | 291 | 11.5 | 25:1 |
| 9 | 19 | 71 | 50 | 100 | 173 | 4.5 | 20 | 0.22:1 |
| 10 | 12 | 88 | 50 | 130 | 178 | 2.5 | 3 | 0.83:1 |
| 11 | 19 | 71 | 50 | 130 | 238 | 8.5 | 17 | 0.5:1 |
| 12 | 25 | 75 | 50 | 130 | 240 | 93 | 5.5 | 17:1 |
| 13 | 25 | 75 | 50 | 180 | 180 | 0 | 0 | - |
| 14 | 12 | 88 | 50 | 180 | 240 | 7 | 5.5 | 1.3:1 |

The sample electrode temperature is 20°C for all recipes.

[a]Proportion of the total gas flow, which was chosen between 34.5 and 37.5 sccm.

[b]Etched for 4 min.